\documentclass[twocolumn,twoside,slac_two]{revtex4}
\usepackage{graphicx}
\usepackage{fancyhdr}
\begin{document}

%Title of paper
\title{Evidence for Shocks as the Origin of Gamma-Ray Flares in Blazars}

\author{M. F. Aller, P. A. Hughes, H. D. Aller}
\affiliation{Astronomy Dept., University of Michigan, Ann Arbor, MI 48109-1042, USA}
\author{A. P. Marscher, S. G. Jorstad}
\affiliation{Institute for Astrophysical Research, Boston University, Boston, MA 02215, USA}

\author{T. Hovatta}
\affiliation{Dept. of Physics, Purdue University, West Lafayette, IN 47907, USA; Cahill Center for Astronomy \& Astrophysics, Caltech, Pasadena, CA 91125, USA}

\author{M. C. Aller}
\affiliation{Dept. of Physics \& Astronomy, University of South Carolina, Columbia, SC 29208, USA}

\begin{abstract}

We present centimeter-band total flux density and linear polarization light curves illustrating the signature
of shocks during radio band outbursts associated in time with $\gamma$-ray flares detected by the {\it Fermi} LAT.
The general characteristics of the spectral evolution during these events is well-explained by new radiative transfer
simulations incorporating propagating oblique shocks and assuming an initially turbulent magnetic field. This finding supports the idea that
oblique shocks in the jet are a viable explanation for activity from the radio to the $\gamma$-ray band in at least some $\gamma$-ray flares.
\end{abstract}

%\maketitle must follow title, authors, abstract
\maketitle

\thispagestyle{fancy}

\section{Overview}
Since the mid-1980s, the leading paradigm for the production of flares in AGN in the radio-to-optical bands has
been shocks which propagate down the relativistic jets of these sources \cite{hug85,mar85}; hydrodynamical simulations have demonstrated
that these structures develop naturally within the jets  \cite{hug05}, making this scenario
a plausible explanation. In contrast, the location and nature of the emission site giving rise to 
the GeV $\gamma$-ray flares, detected first by EGRET and now by the {\it Fermi} LAT,
have remained contentious issues. While both a site near to the central engine and  a location in the parsec scale jet have
been discussed in the literature, mounting evidence based on correlated broadband activity (including high
resolution VLBA imaging of the inner jet) supports a location within the jet
at a site near to the millimeter-band radio core (e.g. \cite{agu11a} and references therein). 
This result supports a direct relation between the flaring in the radio and in the $\gamma$-ray  spectral bands and the production
of the flaring by the same disturbance. The emitting region itself could take the form of a shock (standing or propagating) or a `blob'
with a chaotic magnetic field where turbulence accelerates the high energy electrons. 

The propagating shock scenario, used successfully for the radio band
data has now been proposed as a possible explanation for the $\gamma$-ray flares
in some recent studies, e.g. \cite{abd10}. However, this hypothesis has not been rigorously tested. 
Radiative transfer modeling carried out in the mid-to-late 1980s and early 1990s incorporating
transverse shocks successfully reproduced the spectral evolution of the total flux density and linear
polarization in centimeter-band monitoring data during a few
carefully selected radio band events. However, the transverse shock models failed to match the variability in later events in the same
sources; there the swings in electric vector position angle (hereafter EVPA) were through much less than the 90$^{\circ}$ associated
with transverse structures. This discrepancy indicated that
theoretical explorations incorporating  shocks at other angles to the flow direction
were required. 

To test whether or not propagating shocks play a significant role in the production of the $\gamma$-ray flares, and  ultimately to
set constraints on the physical conditions in the radio jet during $\gamma$-ray flaring, we are 
monitoring the total flux density and linear polarization at 14.5, 8.0, and 4.8 GHz in a sample of 24 blazars
using the University of Michigan 26-m paraboloid (UMRAO), and we are developing
new radiative transfer models allowing for oblique shocks for comparison with these data. Results
are presented below.

\section{ Observational Methodology}
A specific observational goal of our program during
the time period discussed here was to look for the theoretically-expected shock signature 
in the centimeter-band light curves
during $\gamma$-ray flares -- a swing in the EVPA
 and associated flaring in the total flux density and linear polarization (LP).
The changes in LP result from the compression of an initially tangled magnetic field with the
passage of a shock which increases the degree of order. The EVPA, orthogonal to the magnetic field direction
in a transparent source, is a direct measure of this magnetic field orientation in the
emitting region.

The  24 blazars monitored most intensively during the time period discussed here were
3C~66A, 0235+164, 0420-014, 0454-234, 0528+134, 0716+714, 0727-115, 0805-077,
OJ~287, 0906+015, 1156+295, 1222+216, 3C 273, 3C~279, 1329-049, 1502+106,
1510-089, 1633+382, 3C~345, NRAO~530, OT~081, BL~Lac, CTA~102, and 3C~454.3.
These AGN were chosen for detailed study both because they
are bright and variable in the GeV $\gamma$-ray band, and because they have exhibited
well-resolved flares at centimeter-band in historical monitoring measurements. Additionally,
they are all members of the MOJAVE program so that imaging data, useful for disentangling the
flux  contributions from the individual core and jet components, are available at 15 GHz, an overlapping
frequency \cite{lis09}. Eighteen
are members of the Boston University 43 GHz VLBA program (see
http://www.bu.edu/blazars/VLBAproject.html); these monthly imaging data provide important constraints
 on the jet emission properties and in combination with the centimeter band monitoring data give
information on the  opacity between the millimeter core and the 14.5 GHz emission region.

\begin{figure}
%\centering
\includegraphics[width=75mm]{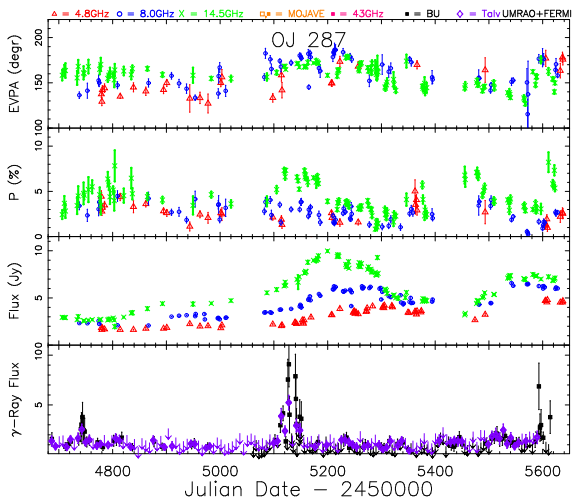}
\caption{Gamma-ray (bottom) and centimeter band light curves (panels 2-4) at 14.5, 8.0.
and 4.8 GHz for the BL~Lac object OJ~287. The bottom panel shows the $\gamma$-ray light curve (E$>$100 MeV) obtained using two
different binning schemes. Squares denote reductions with a bin size adjusted to the variability state (1 and 5 day binning
during flare and non-flare phases respectively). Diamonds denote the light curve generated using 7-day binning throughout. 
The shorter binning of the data during the $\gamma$-ray flare peaks emphasizes the
rapidity of the events in the $\gamma$-ray band and reveals the complex structure which is smoothed out using the 1-week binning,
Units of the $\gamma$-ray light curves are 10$^{-7}$ photons/cm$^2$/sec.
Panels 2 through 4 show from bottom to top daily averages of the total flux density (S), percentage LP, and EVPA. The 
symbols denoting the 3 radio band frequencies are indicated in the upper left.
}
\label{OJ287Rome-f1}
\end{figure}

\begin{figure}
\includegraphics[width=75mm]{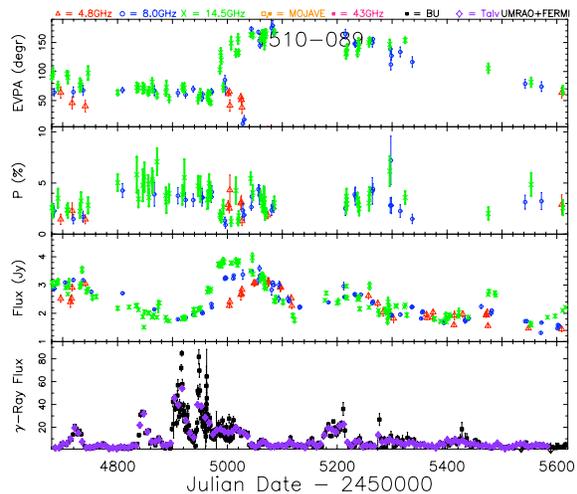}
\caption{ Gamma-ray and centimeter band light curves for the QSO
PKS 1510-089. The bottom panel shows the $\gamma$-ray light curves generated using two different binning schemes.
 Squares denote reductions with
a bin size adjusted to the variability state (3 days, 1 day and intraday). Diamonds denote the light curve
obtained using 7-day binning throughout. The top three panels show
daily averages of the centimeter-band total flux density (panel 2) and the linear polarization
 (panels 3 \& 4). Units and symbols are as in Figure~\ref{OJ287Rome-f1}.}
 \label{1510Rome-f2}
\end{figure}

\begin{figure}
\includegraphics[width=75mm]{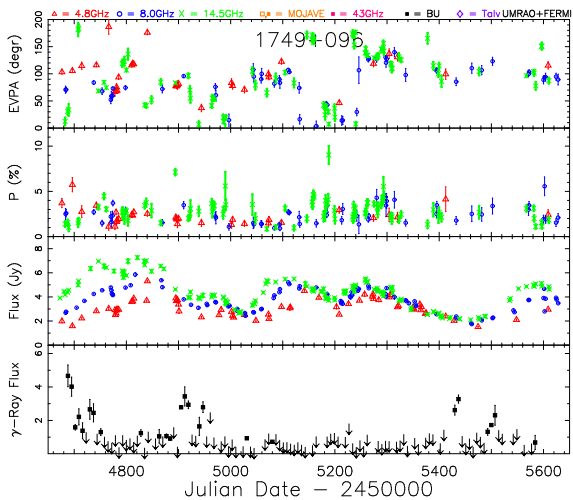}
\caption{Gamma-ray and centimeter band light curves for OT 081 (1749+096) The bottom panel shows the {\it Fermi} light curve
generated using 7-day binning. Daily averages of the radio band total flux density (panel 2) and the linear polarization (panels 3 \& 4)
are shown in the top three panels. Several increases in percentage LP and ordered swings in EVPA can be seen, indicating rapid
and complex radio-band behavior. Units and symbols are as in Figure~\ref{OJ287Rome-f1}.}
\label{1749Rome-f3}
\end{figure}

 The UMRAO observations of the program sources were typically obtained
twice per week at 14.5 GHz and once per week at 8.0 and at 4.8 GHz. This cadence
 was selected to match the expected variability time scales in linear polarization
 and total flux density in the radio band; it was increased, if required, to track the variations 
in more detail during individual flares exhibiting  relatively rapid
flux changes. The sampling rate was highest at 14.5 GHz where the variations
are well-documented to be the highest in amplitude and the most rapid. Each  daily `observation'
consisted of a series of on-off measurements over a 25 to 45 minute time window. These source measurements
were interspersed with observations of positionally-nearby calibrators (selected from a grid) every 1 to 2 hours in order to
 measure the antenna gain and to verify the telescope pointing and  instrumental polarization.

\begin{figure}
\includegraphics[width=95mm]{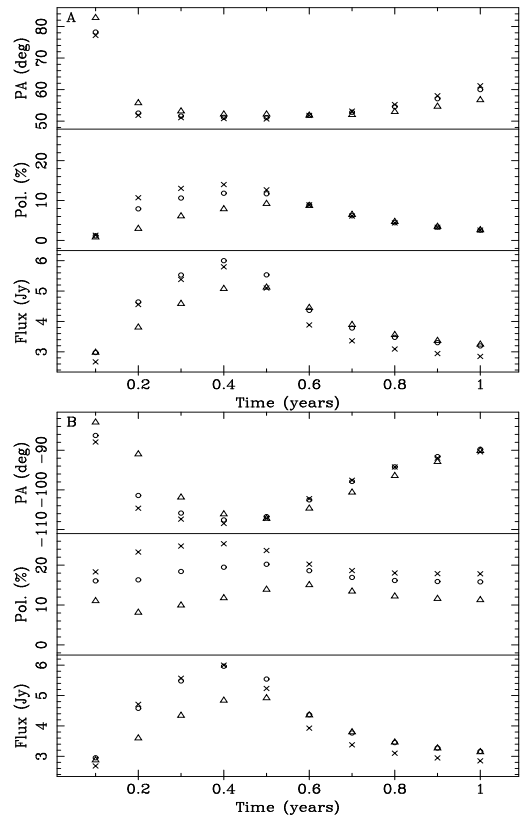}
\caption{Simulated light curves in total flux density and linear polarization
assuming a shock oriented obliquely to
the flow direction, Top (A): The magnetic field is assumed to be purely random. Bottom (B): the
magnetic field is purely ordered (helical).
The symbols correspond to the three UMRAO frequencies shown in Figures 1-3.}
\label{RT-f4}
\end{figure} 

\subsection{Observational Results}
Figures 1 through 3 show examples of the shock signature in the radio band data at or near to times at which large-amplitude
flares were detected in the $\gamma$-ray band by the {\it Fermi} LAT for three of our program sources:
OJ 287, PKS 1510-089, and OT 081. During these events, swings in EVPA through tens of degrees, increases in
fractional LP, and flares in total flux density lasting
for a few weeks to a few months were detected. Time delays and reduced amplitudes are apparent at
the lowest frequency, 4.8 GHz, characteristic of self-absorption.
Specific features to note are: 1) in OJ 287 the two  $\gamma$-ray flares circa JD 2455110 and 2455580 preceded rises in
the fractional LP to levels near 10\% and ordered swings though a limited range of about 40$^{\circ}$ in EVPA.
 The preceding $\gamma$-ray flare near JD 2454740 associated
with a millimeter flare  (see \cite{agu11b}) is not resolved in our data. That study
argued that the flares analyzed were triggered by the interaction of VLBI-scale components with a standing shock;
2) a series of large amplitude events in the QSO PKS 1510-089
beginning circa JD 2454900 preceded a large flare in total flux density and a swing in EVPA. The broadband
behavior is discussed in \cite{mar10} which attributed flaring to the
passage of a knot through a standing conical shock; and 3) complex behavior in the
light curves of the BL~Lac object OT~081 (1749+096). Radio band flares
in this same source during the 1980s were fit with models
incorporating a propagating transverse shock (see \cite{hug91}). 

\section{The Radiative Transfer Modeling}
  Support for the shock model as an explanation for the major outbursts seen in single-dish data,
  and the propagating components seen in maps of parsec-scale flows, as well as
  support for a shock explanation for at least some {\it Fermi} events,
  would come from `revalidating' the `shock in jet' model, by showing
  that oblique shocks can indeed explain the commonly observed reduced
  swing in EVPA through only tens of degrees, and associated increases
  in both percentage LP and total flux density (flares) with
  the spectral behavior exhibited in the data.
To quantitively test whether the shock scenario can reproduce the features in the
data, we have developed radiative transfer models which allow for a shock at
 any orientation relative to the flow direction. The models discussed here do not allow for retarded
time effects which will be included in future modeling. The shock is assumed to propagate at a constant rate (no acceleration or
deceleration). A detailed description of the formulation is  presented in \cite{hug11}.
Representative simulations illustrating the characteristics of
the simulated spectral evolution are shown in Figure~\ref{RT-f4}. 
Those presented  assumed
a compression of 0.7, a forward moving shock, a shock obliquity of 45$^{\circ}$, and a viewing angle of
10$^{\circ}$.  As a test of the effect of the magnetic field degree of order on the
emission properties and to better understand the jet conditions, light curves were generated assuming a
purely ordered (helical) magnetic field, a purely random magnetic field, and a
mix of the two. Results are shown for two of these cases: a purely ordered (bottom) and
a turbulent (top) magnetic field. Comparison with the data shows that the spectral
evolution for the helical case does not match either the observed maximum 
fractional LP (the predicted maximum fractional LP $>$ 20\% at 14.5 GHz exceeds the maximum observed value in
the UMRAO database of LP observations) or the observed spectral evolution in EVPA. A purely
turbulent ambient magnetic field yields the best fit with the data.

\section {Summary of Results}
 The expected shock signature was successfully identified in the radio band data during several {\it Fermi}-detected
events; this result is  consistent with a shock-in-jet origin for at least some $\gamma$-ray flares.
The ordered swings in EVPA typically occurred on timescales of weeks-to-months over a range of tens of degrees. These
characteristics and the spectral evolution in linear polarization and total flux density are
explained well by our theoretical simulations which incorporate realistic flow conditions. 
Specific features of the data  during radio band
outbursts which are reproduced by the new models are: the fractional total flux density increase,
 the spectral evolution in total flux density through a partially optically thick phase
during outburst rise, the magnitude of the peak percentage
polarization with opacity/Faraday effects evident at the lowest frequency, and a swing in EVPA through tens
of degrees. The effect of differences in the
degree of order on the emission have been investigated for the cases of a  purely turbulent magnetic
field, a purely ordered magnetic field, and a mix of the two. Comparison with the
data shows that the observed characteristics are best explained
by a model in which the magnetic field within the density enhancement is predominantly random before it
passes through the shock.

 While detailed light curves were not computed for the
standing shock case, we expect that the general characteristics of
these light curves will be similar to those for the
propagating shock discussed here. The structure of such a feature
is unlikely to be simple; at best it will be biconical, and possibly have a Mach stem, and an oblique
shock will capture some, but not all, of the attributes of a disturbance
passing through the conical structure. 
Observing and modeling repeated events
from the {\it same} source will be an important next  step in pinning down the correct hydrodynamic description of
the jets studied and in identifying the origin of the $\gamma$-ray emission in specific events. 

\bigskip % extra skip inserted
\begin{acknowledgments}
This research was funded by NASA grants NNX09AU16G and NNX10AP16G and by NSF grant
AST-0607523 (University of Michigan), NASA grants NNX08AV65G, NNX08AV61G, NNX08AJ64G, NNX09AT66G, and
NNX10AU15G, and NSF grant AST-0907893 (Boston University), and NASA 
grant NNX08AV67G and NSF grant AST-0807860 (Purdue University). Funding for the operation of UMRAO was provided by
the University of Michigan.
\end{acknowledgments}
\bigskip % extra skip inserted
%\begin{thebibliography}{9}   % Use for  1-9  references

\end{document}